\documentclass[twocolumn,pre,amsmath,amssymb,floatfix,aps,superscriptaddress]{revtex4}
\usepackage{subfigure}
\usepackage{color}

\usepackage{graphicx}
\usepackage{bm}  
\usepackage{epsfig}
\usepackage{mathtools}

\begin{document}

\title{Neutral versus charged defect patterns in curved crystals}

\date{\today}

\author{Amir Azadi}
\affiliation{Department of Physics, Harvard University, Cambridge, MA 02138, USA}

\author{Gregory M. Grason}
\affiliation{Department of Polymer Science and Engineering, University of Massachusetts, Amherst, MA 01003, USA}

\begin{abstract}
Characterizing the complex spectrum of topological defects in ground states of curved crystals is a long-standing problem with wide implications, from the mathematical Thomson problem to diverse physical realizations, including fullerenes and particle-coated droplets.  While the excess number of ``topologically-charged" 5-fold disclinations in a closed, spherical crystal is fixed, here, we study the elementary transition from defect-free, flat crystals to curved-crystals possessing an excess of ``charged" disclinations in their bulk.  Specifically, we consider the impact of topologically-neutral patterns of defects -- in the form of multi-dislocation chains or ``scars" stable for small lattice spacing -- on the transition from neutral to charged ground-state patterns of a crystalline cap bound to a spherical surface.  Based on the asymptotic theory of caps in continuum limit of vanishing lattice spacing, we derive the morphological phase diagram of ground state defect patterns, spanned by surface coverage of the sphere and forces at the cap edge.  For the singular limit of zero edge forces, we find that scars reduce (by half) the threshold surface coverage for excess disclinations.  Even more significant, scars flatten the geometric dependence of excess disinclination number on Gaussian curvature, leading to a transition between stable ``charged" and ``neutral" patterns that is, instead, critically sensitive to the compressive vs. tensile nature of boundary forces on the cap. 
\end{abstract}

\maketitle


\section{Introduction}
\label{intro}

\begin{figure}
\center\includegraphics[width=0.3\textwidth]{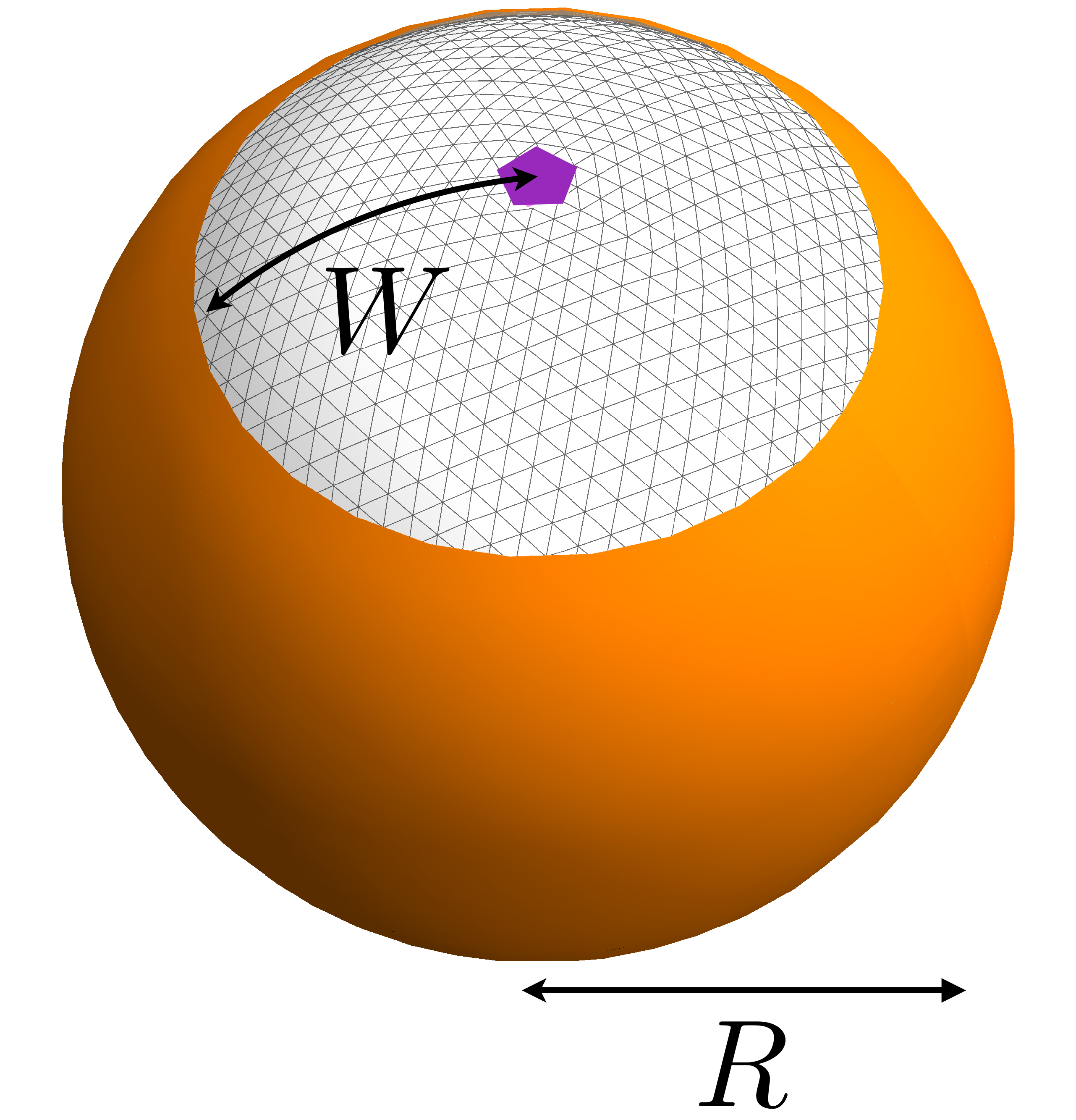}
\caption {The model geometry: a 2D circular crystal, radius $W$, formed on a (rigid) spherical surface of radius $R$.  A 5-fold disclination (purple pentagon) is shown at the cap center.}
\label{Fig1}
\end{figure}

Optimizing ground state order, typically a trivial affair on planar surfaces, becomes one of a number of unsolved, century-old problems purely through the introduction of positive Gaussian curvature, arguably the simplest (i.e. most isotropic) non-flat geometry \cite{smale}.  Motivating the goal to understand optimal ordering on spheres, known broadly as the generalized-Thomson problem, is structure formation in material systems as diverse as viral capsids \cite{caspar,bruinsma_pnas}, fullerenes \cite{ful}, particle-coated droplets \cite{bausch, irvine}, curved bubble rafts~\cite{shin}, emulsion droplets ~\cite{sloutskin} and spherical superconductors  \cite{sym_thomson}.  Ground states in these systems are characterized by topological defects, 5- and 7-fold disclinations in otherwise sixfold hexagonal packing, which carry, respectively, positive and negative topological  charges $s_i=\pm  \pi/3$ associated with the rotation of lattice directions around the defect~\cite{nelson_book}.  The preponderance of studies that have addressed this problem consider closed spherical shells, where the Gauss Bonnet theorem requires a fixed topological charge $S \equiv \sum_i s_i = 4 \pi$, or exactly, twelve more 5-fold than 7-fold defects~\cite{altschuler, bowick, bowick_caccuito, wales_09, bausch}.  Far less understood, and the subject this article, is the progression of order from neutral ($S=0$), flat crystals to topologically-charged ($S>0$) bulk order as surface curvature is increased.  In this article, we address this transition for a ``crystalline cap" covering an incomplete fraction, $\Phi$, of the sphere (see Fig.~\ref{Fig1}).

The free boundary of the crystal introduces several complicating factors in determining ground state.  Foremost among these is possibility of external forces acting at the boundary, which may be tensile, as in the case of adhesion between a crystal and spherical substrate, or compressive, for say a crystalline droplet pinned at the edge.  How external stresses interact with those imposed by geometry in the stabilization of lattice defects is an open question. Even more significant is the fact that free boundaries allow the net charge of disclinations in the interior to vary according to energetic considerations, rather than topological constraints.  Not only can the {\it net} charge of defects vary with increasing curvature, but so too can the number of {\it neutral} 5-7 pairs, or dislocations, in curved crystals~\cite{irvine}.  Extended chains, or ``charged" dislocation scars, where first predicted and observed in closed spherical crystals of sufficient particle number~\cite{moore, bowick}, where they extend from isolated 5-fold disclinations, growing in number and pattern complexity (e.g. multi-arm scars and rosettes) as number of lattice points grows large~\cite{wales_09}.  Dislocation chains, or neutral scars (possessing equal numbers of 5- and 7-fold defects), are also also stable in weakly-curved crystals possessing no excess disclinations ($S=0$)~\cite{azadi_prl, wales_13} and experiments on negatively-curved colloidal crystals by Irvine and coworkers further suggest the formation or neutral scars, or ``pleats", may delay the onset of excess disclinations in curved crystals~\cite{irvine}.  Precisely how the large numbers of dislocations alters the relative stability of charged vs. neutral defect states of curved crystals remains unknown.

In this article, we explore the transition from neutral- to charged-defect ground states for crystalline caps on spherical substrates subject to external boundary forces, $\sigma_b$.  Our analysis is based on the continuum elasticity theory of curved 2D crystals and exploits, in particular, a newly developed asymptotic theory of multi-dislocation crystals~\cite{davidovitch_grason, azadi_prl} predicated on the principles of ``optimal stress collapse" in scarred zones of the crystal.  Beyond the parameters $\Phi$ and $\sigma_b$, the structure and energetics of crystals is also critically sensitive to the ratio $b/W$, where $b$ and $W$ are respective lattice spacing and lateral crystal size.  The number of dislocations in curved-crystal ground states diverges as $W/b$ in the {\it continuum limit}, the singular limit captured by this asymptotic theory.  Previously, we studied the principles governing the pattern selection of neutral ($S=0$) scar patterns of weakly-curved caps~\cite{azadi_prl}, shown~\cite{davidovitch_grason} to be formally analogous to the wavelength selection of in optimal wrinkle patterns in radially-confined thin sheets~\cite{schroll, king}.  In the present study, we exploit this asymptotic theory for crystalline membranes to determine the multi-dislocation patterns that achieve an optimal level of stress relaxation for both neutral caps and those possessing a single excess disclinations, and show, by comparison of the dominant energetics in the $b/W \to 0$ limit, that excess dislocations have a profound on the relative stability of neutral vs. charged defect patterns in caps.

While, in the absence of excess dislocations, the critical curvature separating neutral and charged patterns has a simple linear dependence on $\sigma_b$, as $b/W \to 0$ and dislocations abound, multi-scar patterns are shown to profoundly ``flatten" the dependence of this boundary on geometry (or $\Phi$).  This leads to a neutral vs. charged morphology phase diagram that, in this asymptotic limitm is entirely governed by the tensile ($\sigma_b <0$) vs. compressive ($\sigma_b>0$) nature of boundary forces.  Hence, we find that the case of zero boundary forces is a singular limit, where the stability of neutral vs. charged patterns is determined by the lowest-order corrections to the energy for small, but finite $b/W$.  We show by combination of scaling analysis and numerical simulations, that the dependence of these sub-dominant energetics on $\Phi$ (or equivalently, curvature) can largely be attributed to the variation of number of defects needed to perfectly relax geometrically imposed stresses.  Excess dislocations significantly shift the transition to charged defect patterns to lower curvature:  from $\Phi= 1/6$ in the absence of dislocations to $\Phi \simeq 1/12$ when dislocation scars proliferate as $W/b \to \infty$.

The remainder of this manuscript is organized as follows.  In Sec. \ref{model}, we introduce the continuum model of spherical caps and analyze the transition from neutral to charged caps in the absence of excess dislocations.  In Sec. \ref{collapse} we introduce the principles underlying the structure of multi-dislocation scar patterns in the $b/W\to0$ continuum limit.  In Sec. \ref{compete} we summarize the competing patterns of stable charged and neutral scars predicted by this theory and the ``defect phase diagram" for the  $b/W\to0$ continuum limit in Sec. \ref{continuum}.  In Sec. \ref{finiteep}, we consider the singular limit of vanishing boundary forces ($\sigma_b=0$), and show how the finite-$b/W$ corrections to the energy along this line govern the transition from neutral to charged caps with increasing curvature.  Finally, we conclude with discussion of our results and remaining open questions regarding defect patterns in curvature-frustrated crystals.

\section{Model and Dislocation-Free Limit}
\label{model}
We consider a crystalline domain bound to a sphere of fixed radius, $R$ (Fig.~\ref{Fig1}).  We assume the domain has circular shape of radius $W$ (e.g. no edge faceting), such that the {\it surface coverage} is simply
\begin{equation}
\Phi = \Big(\frac{W}{2R} \Big)^2
\end{equation}.  We employ a continuum elastic theory to model the deformation of hexagonal order in the domains, described by the energy 
\begin{equation}
\label{eq: energy}
E = \frac{1}{2} \int dA ~ \sigma_{ij} u_{ij} - \sigma_b( \Delta A) ,
\end{equation}
where $u_{ij}$ and $\sigma_{ij}$ are the respective in-plane elastic strain and stress tensors, satisfying the linear constitutive relation $u_{ij}=Y^{-1} \big[ (1+\nu)\sigma_{ij} - \nu  \delta_{ij} \sigma_{kk} \big] $ where $Y$ and $\nu$ are the 2D modulus of Poisson ratios of the hexagonal crystal.  The second term represents the work done an boundary force $\sigma_b$ to change the area of the crystal by $\Delta A=W \int d\theta~u_r(r=W)$, where $r$ and $\theta$ are polar coordinates.  Depending on the physical realization of the crystalline cap, boundary forces $\sigma_{b}$ may be either tensile (e.g. due to adhesive spreading on a substrates) or compressive (e.g. due to pinning at cap edge).

\begin{figure}
\center\includegraphics[width=0.4\textwidth]{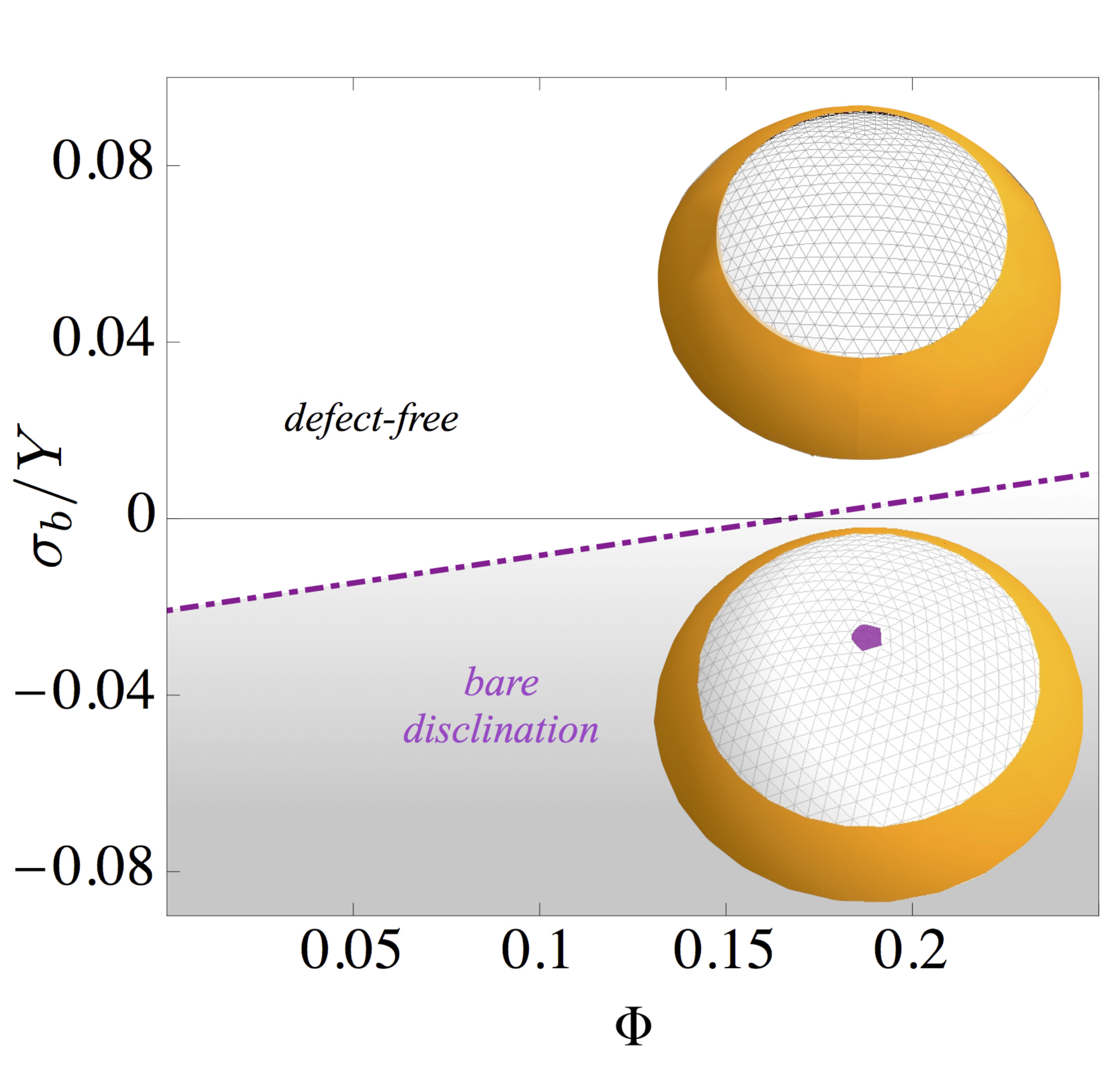}
\caption {Phase diagram spanned by $\Phi$ and external boundary stresses, $\sigma_{b}/Y$ for crystalline caps without excess disclinations, relevant to lattice spacings comparable to cap width, $b \lesssim W$.}
\label{Fig2}
\end{figure}

To understand the conditions where the first excess disclination becomes stable, we restrict our focus to the nearly-flat caps, $\Phi \ll 1$, where the elastic stress of the crystal may be calculated from the F\"{o}ppl- von K\'{a}rm\'{a}n theory of weakly-deflected, crystalline membranes \cite{nelson_seung, nelson_book}.  The stress is determined by in-plane force balance $\partial_i \sigma_{ij} = 0$, boundary conditions $\sigma_{rr}(r=W) = \sigma_b$ and the compatibility condition,
\begin{equation}
\label{com_5fold}
Y^{-1} \nabla^2_\perp \sigma_{ii} = -R^{-2} + s \delta ^{2}(\mathbf{x})+\nabla_\perp \times {\bf b} ({\bf x}) .
\end{equation}
The right-hand side of this equation describes intrinsic sources of stress in the sheet, including the imposed Gaussian curvature, $R^{-2}$, and two types of topological defects, disclinations and dislocations.  The monopole source represents the possibility of disclination of charge $s$ centered at ${\bf x}=0$:   $s=0$ for neutral caps and $s = +  \pi/3$ for charged caps (a 5-fold defect)~\footnote{In this continuum theory, we do not consider the presence of ``disclinations" at the boundary of the hexagonal crystal~\cite{giomi_bowick} associated with deficit neighbors at the facet corners, as these generate no elastic distortion (or stress) in the crystal.}  The third term derives from the Burger's density of dislocations ${\bf b} ({\bf x}) = \sum_\alpha  {\bf b}_\alpha \delta({\bf x} - {\bf x}_\alpha)$, where $ {\bf b}_\alpha $ is the Burger's vector of a dislocation at ${\bf x}_\alpha$.  Dislocations are neutral 5-7 disclination dipoles, and like the polarization charge in electrostatics, gradients~\footnote{Here,$\nabla_\perp \times {\bf b}  = \epsilon_{ij} \partial_i b_j$ is the 2D curl.} of ${\bf b} ({\bf x})$ generate sources of elastic stress~\cite{nelson_book}.

A key challenge to comparing the structure and stability of competing defect patterns is the determination of ${\bf b} ({\bf x})$, as dislocations vary in location, orientation and number according to their ability relax elastic stress.  Below we describe the governing principles and implications of the optimal dislocation distributions when dislocations are numerous.  In this section, we first consider dislocation-free caps, which arise when self-energy of dislocations exceeds their relaxation of geometric stresses, as quantified by the dimensionless ratio of the self-energy of dislocations ($\propto Y b^2)$ and the elastic energy of geometric confinement ($\propto Y \Phi^2 W^2)$,
\begin{equation}
\epsilon = (b/W)^2 \Phi^{-2} 
\end{equation}
Caps are dislocation free for $\epsilon \gtrsim 1$, as is the case when crystal widths are comparable to lattice spacing, $W \gtrsim b$.  Taking ${\bf b} ({\bf x})=0$ and the axisymmetric stress of the dislocation free cap has the general form, 
\begin{eqnarray}
\label{stress_5fold}
\sigma^{\rm df}_{rr}&=&\frac{Ys}{4\pi}\ln (r/W)+\frac{Y\Phi}{4} \Big[1 - \Big(\frac{r}{W}\Big)^2\Big]+\sigma _{b};\nonumber\\  \sigma^{\rm df}_{\theta\theta}&=&\frac{Ys}{4\pi}[\ln (r/W)+1]+\frac{Y\Phi}{4} \Big[1 - 3\Big(\frac{r}{W}\Big)^2\Big] + \sigma _{b}\nonumber\\ 
\end{eqnarray} 
and $\sigma^{\rm df}_{r \theta}=0$.    The first terms in eq. (\ref{stress_5fold}) are related to the disclination induced stresses, with $s=+ \pi/3$ for five-fold defects and $s=0$ for the neutral state. The The second contributions derive from curvature-induced strains, which become increasingly compressive with distance from the center.  

Solving for the strain and displacement fields from $\sigma_{ij}^{\rm df}$ and inserting into main text eq. (\ref{eq: energy}) we have the total energy of the dislocation-free state,
\begin{equation}
\label{eq: energydf}
\frac{E_{\rm df} (s) }{Y \pi W^2}=\frac{  \Phi^2}{24} -\frac{(1-\nu)\sigma_{b}^2}{Y^2} +\frac{\Phi \sigma_b}{Y}
 +\frac{s^2}{32 \pi^2}-\frac{ s}{2 \pi}\Big(\frac{\Phi}{8} - \frac{\sigma_{b}}{ Y}\Big)
 \end{equation}
 While the first three terms represent the mechanical energy of imposed geometric and boundary stresses, the final terms represent the disclination self-energy and the mechanical coupling between the disclination and geometric and external stresses.  Due to the far-field tensile nature of five-fold defects, the cost of positive disclinations decreases with increasing external or geometrically-induced compression.
 
Comparing the energy of neutral $(s=0)$ to charged $(s = + \pi /3)$ caps, we find the critical surface coverage $\Phi^*_{\rm df}$ above which dislocation-free caps favor an excess disclination (plotted in Fig.~\ref{Fig2}),
\begin{equation}
\Phi^*_{\rm df} (\sigma_b) =\frac{1}{6} +\frac{\sigma_{b}}{8 Y} .
\end{equation}
Due to the imperfect screening of the geometric stresses by singular disclinations, in the absence of external forces the integrated Gaussian curvature must exceed the point at which it ``neutralizes" the topological charge (i.e. $\Phi =1/12$) of the first stable disclination.  In terms of integrated Gaussian curvature, this threshold corresponds to $2 \pi/3$, which is comparable to the threshold curvature ($\simeq 2.5$)  when a single 5-fold disclination becomes stable in a parabolic crystal, which is also more than double what is needed to compensate for the deficit angle of the disclination~\cite{giomi}.  Further, due to the far-field tension generated by disclinations, boundary compression ($\sigma_b<0$) enhances stability, such that they become stable in even in {\it flat caps} when $\sigma_b \leq -4 Y/3$.

\begin{figure*}
\center\includegraphics[width=0.95\textwidth]{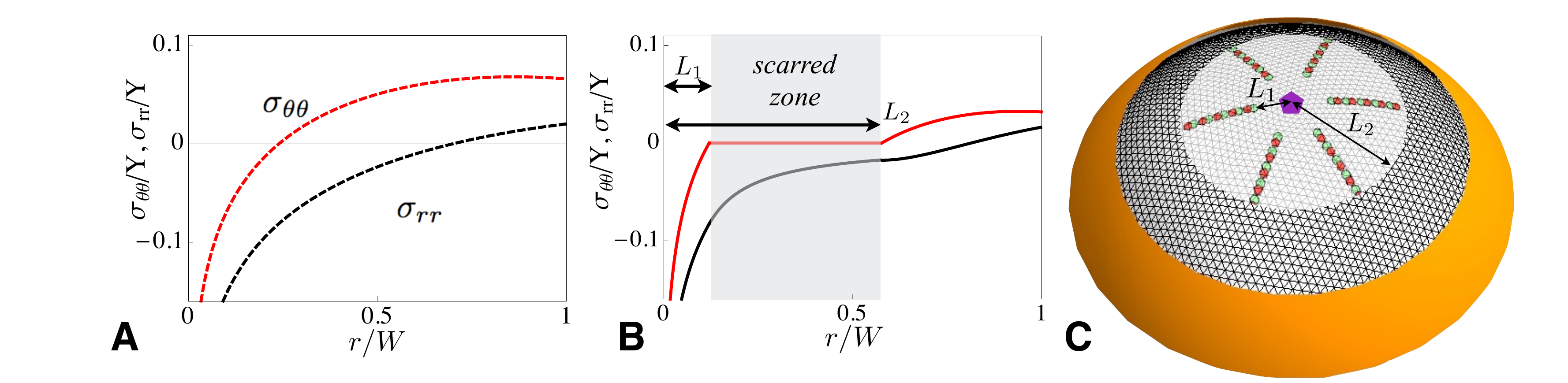}
\caption {(A) Example profiles for dislocation-free stress fields for a cap with a centered 5-fold disclination at $\Phi=0.07$ and $\sigma_b = 0.02 Y$.  Hoop and radial stress components are shown as red and black curves, respectively.  (B) Example profiles for stress-collapsed mechanical state for same conditions as (A), possessing radial scars in an annular region $L_1\leq r \leq L_2$, correspond to the schematic of the ``charged scar" pattern in (C), where scars are shown as alternating 5-7 pairs (red pentagons and green heptagons, respectively).  }
\label{fig3}
\end{figure*}

\section{Principles of multi-dislocation stress collapse}
\label{collapse}

In this section, we outline the principles underlying optimal multi-dislocation patterns, in the asymptotic limit $\epsilon \to 0$, where the lattice spacing vanishes with respect to $W$, and the energetic costs of dislocation vanish in comparison to the costs of geometric frustration.  In this limit the intra-cap stress is profoundly modified from the dislocation-free case described above by large numbers of dislocations which we assume to constitute multi-dislocation scars.   To construct stable patterns of stress in the $b \to 0$ continuum limit, we note that the elastic force on a dislocation ${\bf b}$ is proportional to the total stress as $\epsilon_{ij} \sigma_{jk} b_k$.~\cite{peach}  Comparing this to the stabilizing pull of ``self-interactions" of dislocations that drive it towards the boundary and that scale as $b^2$, it is clear that in the $b\to0$ limit components of stress along the Burgers vector orientation of dislocations must vanish in order that dislocations are under zero force.  To construct a distribution of dislocations satisfying mechanical equilibrium  in axisymmetric caps, we assume their polarization to be along the hoop direction, of ${\bf b} ({\bf x} )= b \rho({\bf x} ) \hat{\theta}$, where $\rho({\bf x} )$ is the areal dislocation density.  This defect polarization is consistent with partial rows of lattice sites added or removed along the radial direction of the crystal, which facilities the collapse of compressive or tensile hoop stresses depending on the sign.  Therefore, we seek stress distributions which satisfy perfect stress collapse along the hoop direction (i.e. $\sigma_{\theta \theta}$) in regions occupied by (stable) dislocations.  

This ``stress-collapse" principle is formally equivalent to the notion of perfect ``screening" of Gaussian curvature,proposed previously for scars on curved crystals~\cite{bowick, irvine}.  Here we argue that this principle is more than a heuristic argument, that it becomes quantitatively accurate in the asymptotic limit $b/W \to 0$, provided that it is generalized to account for the presence of other stresses in the system (e.g. boundary forces).    Outside of the context of strictly crystalline membranes, this stress collapse principle is the exact analogy to the ``tension field", or ``far-from threshold", limit exploited in the analysis of thin sheen wrinkling, where the underlying pattern of elastic stress supports zero compression in the limit of vanishing thickness \cite{king, schroll}.

For both neutral caps and charged caps with centered disclinations, force-free multi-dislocation patterns can be described in terms of the following 3-zone solution (see Fig.~\ref{fig3}):  {\it zone I}, $0<r<L_1$ with or without a central disclination and dislocation-free; {\it zone II} $L_1\leq r <L_2$ an annulus possessing azimuthally-orientated dislocations with density $\rho(r) \neq 0$; and {\it zone III}, $L_2 \leq r \leq W$, the outer dislocation-free zone of the cap.  For the inner zone $0<r<L_1$ (zone I) the stress pattern is identical to the dislocation free cap, up to an additive constant, which guarantees hoop stress to vanish at the edge of the scarred zone, at $r=L_1$,
\begin{equation}
\frac{\sigma^I_{\theta \theta}}{Y}= \frac{s}{4 \pi} \ln (r/L_1) + \frac{3(L_1^2-r^2)}{16 R^2}; \ \frac{\sigma^I_{rr}}{Y} = \frac{\sigma^I_{\theta \theta}}{Y}-\frac{s}{4 \pi}+\frac{r^2}{8R^2} .
\label{zone1}
\end{equation}
For the scarred annulus $L_1\leq r <L_2$ (zone II), mechanical equilibrium for present dislocations requires vanishing hoop stress.  Thus, using radial force balance $\partial_r(r \sigma^{II}_{rr} ) = \sigma^{II}_{\theta \theta}=0$ and matching radial stress at the inner zone edge we find 
\begin{equation}
\frac{\sigma^{II}_{\theta \theta}}{Y}=0; \ \frac{\sigma^{II}_{rr}}{Y} = \Big(\frac{L_1^2}{8 R^2}- \frac{s}{4 \pi} \Big) \frac{L_1}{r} .
\label{zone2}
\end{equation}
Finally, matching radial and hoop stresses at the boundary between the scarred zone at the dislocation-free outer zone ($r=L_2$) with the we find the stress in zone III for $L_2 \leq r \leq W$,
\begin{eqnarray}
\label{zone3}
\frac{\sigma^{III}_{\theta \theta}}{Y} &\!\!=\!\!& \frac{s}{4 \pi} \ln (r/L_2) + \frac{3(L_2^2-r^2)}{16 R^2} - B\Big(\frac{1}{r^2} -\frac{1}{L_2^2}\Big) ;  \nonumber \\ \frac{\sigma^{III}_{rr}}{Y} &\!\!=\!\!& \frac{\sigma^{III}_{\theta \theta}}{Y}-\frac{s}{4 \pi}+\frac{r^2}{8R^2} +\frac{2 B}{r^2} ,
\end{eqnarray}
where $B/L_2 = (s/8\pi) (L_2-L_1) - (L_2^3-L_1^3)/16R^2$.~\footnote{In addition to stress continuity, the solution must guarantee continuity of the first integral of the compatibility equation, eq. \ref{com_5fold}, which requires the coefficient of the logarithmic terms in eq. \ref{zone3} to match bare charge of the central disinclination.}  Finally, requiring that radial stresses reach the imposed boundary force at the edge $\sigma^{III}_{rr} (R=W) = \sigma_b$, we arrive at an ``equation of state" for the scarred caps related to the positions of the scarred zone edge, $L_1$ and $L_2$, to the boundary forces, central disclination charge and to surface curvature,
\begin{multline}
\label{eq: L1L2}
\frac{\sigma_b }{Y} =\frac{L_1}{L_2}\Big(\frac{L_1^2}{16R^2} - \frac{s}{8 \pi} \Big) \frac{(W^2+L_2^2)}{W^2} \\ -\frac{s}{4 \pi} \Big[ \ln (L_2/W)+\frac{W^2-L_2^2}{2W^2}\Big] - \frac{(W^2-L_2^2)^2}{16 W^2 R^2} .
\end{multline}
This relationship governs the gross structure and dominant energetics of competing multi-dislocation patterns.  The real solutions for scar edges satisfying $0\leq L_1 \leq L_2 \leq W$ correspond to patterns of dislocation scars that achieve mechanical equilibrium in the limit of vanishing lattice spacing.  

\begin{figure}
\center\includegraphics[width=0.4\textwidth]{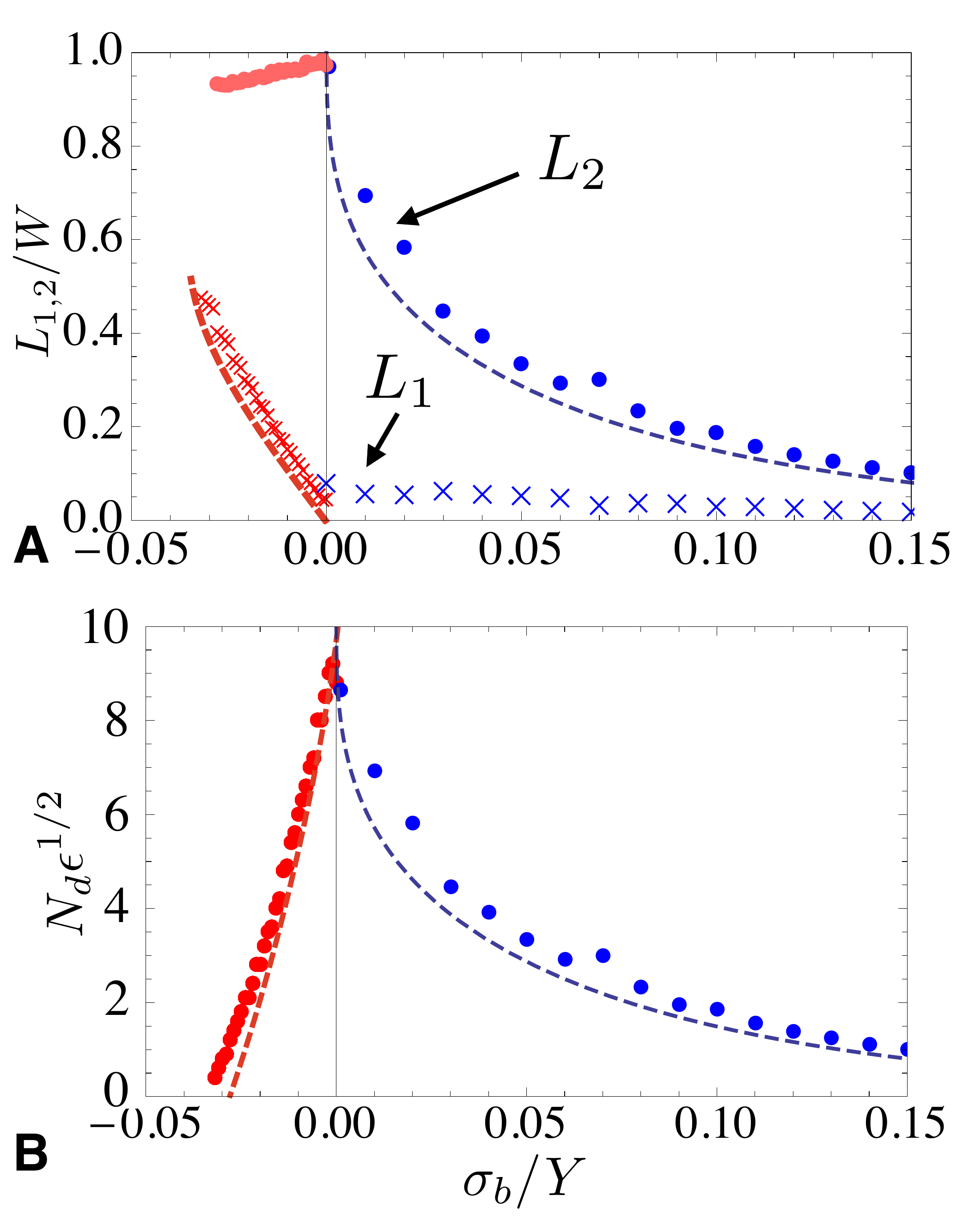}
\caption{(A) shows $n$-fold symmetric scar energy minimization simulation results (points and crosses) for the scar ends, $L_2$ (filled circles) and $L_{1}$ (crosses) for a charged cap ($s=\pi/3$) at surface coverage $\Phi=0.075$, where predictions for $L_1> 0$ ($\sigma_b <0$) and $L_2<W$ ($\sigma_b>0$) from the equations of stress collapse are shown as dashed lines.  (B) shows the total number of dislocation $N_{d}$ scaled by the reduced cap radius, $W/b$, compared to predictions given by eq. (\ref{eq: Ndedge}) and (\ref{eq: Ndcenter}).}
\label{Fig4}
\end{figure}

Given a solution to this eq.~(\ref{eq: L1L2}) for the boundaries of the scarred zone, the dislocation density that achieves the maximal state of stress collapse in the $\epsilon \to 0$ limit follows directly from the integration of the compatibility equation, eq. (\ref{com_5fold}), $2 \pi ( r b_\theta)\big|^r_{L_1} =  2\pi Y^{-1} (r \partial_r \sigma^{II}_{ii}) \big|^r_{L_1}-\pi (r^2-L_1^2)/R^2$ or 
\begin{equation}
b_\theta(r) = \frac{r}{2 R^2}+\frac{L_1}{ r^2} \Big(\frac{ L_1^2}{ 8R^2} -\frac{s}{4 \pi}\Big) -\frac{2 s}{\pi r}, \ \ {\rm for} \ L_1\leq r \leq L_2
\label{br}
\end{equation}
We note the respective positive and negative signs of terms proportional curvature and curvature, indicating that curvature- and disclination-induced stress favor dislocations of opposing polarization in caps.

\section{Neutral- and charged-cap morphologies}

\label{compete}

Solutions to eqs. ~(\ref{eq: L1L2}) and (\ref{br}) govern the multi-scar morphologies of both neutral and charged caps in the continuum limit of vanishingly small lattice spacing.  While eq. (\ref{eq: L1L2}) is a single equation relating the two unknown positions of scar-zone edges, for low surface coverages  solutions that maximize the elastic energy relaxation of both neutral and charged caps fall into one of two states:  center-bound scars, where $L_1 \to 0$ and $L_2\leq W$, and edge-bound scars, where $L_1 \geq 0$ and $L_2 \to W$.  Heuristically, maximizing the degree of energy relaxation by dislocations can be understood as maximizing the width of the stress collapsed zone $L_2-L_1$.  It can be shown, for example for neutral caps, from eq. (\ref{eq: L1L2}) that this corresponds to scars extending to the free boundary of caps ($L_2 \to W$) for tensile boundary forces, while for compression, scars are bound to the cap center ($L_1=0$).  

As more critical test,  we compare the optimal stress-collapse states to discrete-dislocation numerical simulations of performed by relaxing the number and position of dipolar sources of stress in the elastic energy (see Appendix~\ref{simulations} for details).  In these simulations, interactions between dislocations and other sources of stress (curvature, boundary forces and other defects)~\cite{azadi_grason_12} and number and position of dislocations is relaxed numerically assuming an $n$-fold symmetric pattern of radial scars (optimizing with respect to scar number)~\cite{azadi_grason_12}.  In Fig. \ref{Fig4}, we plot results for charged ($s= \pi/3$) caps, position of scar ends, $L_1$ and $L_2$, and total dislocation number, $N_d$, for simulated caps with $W/b = 0.0025$ at surface coverage $\Phi = 0.075$ as function of $\sigma_b$.   Simulation results for charged caps show remarkable quantitative agreement with the respective $L_2 \to W$ and $L_1 \to 0$ solutions of eq. \ref{eq: L1L2} and the total dislocation number $N_d$, derives from the integrate of this distribution over the cap $N_d=b^{-1} \int dA |b_\theta(r)|$.   For sufficiently small curvatures, the polarization of dislocations is constant throughout scars (either clockwise or counter-clockwise), and number of dislocations in edge-bound scars this has the form
\begin{multline}
\label{eq: Ndedge}
N_d(L_2=W) = \frac{2 \pi}{\epsilon^{1/2}} \bigg| -\frac{ 2}{3} \Big( 1- \frac{L_1^3}{W^3} \Big) +\frac{ \sigma_b}{Y \Phi } \ln \Big(\frac{W}{L_1}\Big) \\ + \frac{ s}{2 \pi  \Phi} \Big(1-\frac{L_1}{W} \Big) \bigg|, 
\end{multline} 
while for center-bound scars we find
\begin{equation}
\label{eq: Ndcenter}
N_d(L_1=0) =  \frac{2 \pi}{\epsilon^{1/2}}\bigg|- \frac{ 2}{3}  \frac{L_2^3}{W^3}  + \frac{ s}{2 \pi \Phi} \frac{L_2}{W} \bigg|.
\end{equation}
In both states, the number of dislocations diverges in the continuum limit as $\epsilon^{-1/2}$, asymptoticly approaching the continuous distribution given by $b_\theta(r)$.

\begin{figure*}
\center\includegraphics[width=1.0\textwidth]{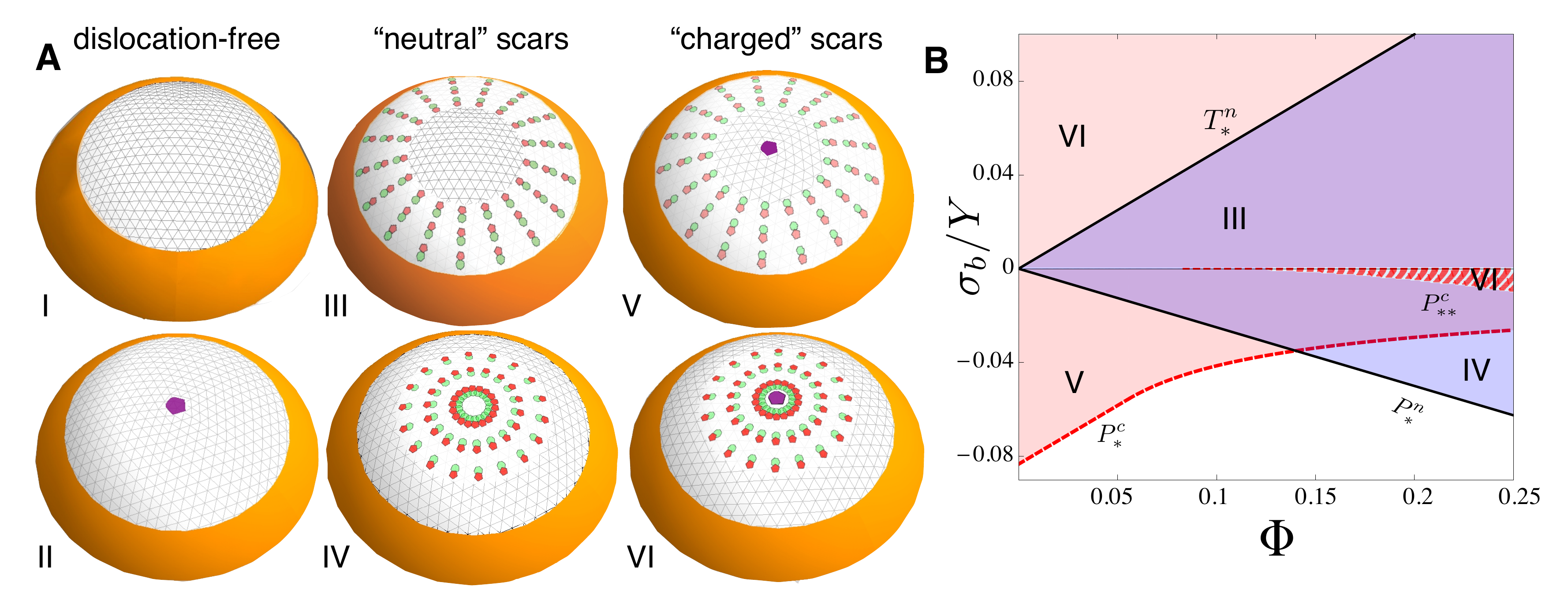}
\caption{A) shows the schematics of six possible states (defect morphologies I-VI) describe in the text.  (B) depicts the phase space spanned by surface coverage, $\Phi$ and boundary stresses $\sigma _{b}$, and delineates the regions of mechanical stability (existence) for the competing radial scar morphologies. The blue region outlines where neutral scars  exist (states III or IV), bounded by two critical solid lines $\sigma<T^n_{*}$ and $\sigma>-P^n_{*}$. The red region corresponds to the stable charged caps (states V or VI), bounded from below by critical curve $\sigma>-P^c_{*}$.  The hashed red region (bounded by  $\sigma>-P^c_{**}$) outlines the existence of center-bound, charged caps (state V) for compressive boundary forces. Dislocation-free states (I \& II) exist for all curvatures and boundary forces.}
\label{fig5}
\end{figure*}

Based on the condition of edge-bound or center-bound scar states, six distinct competing ``defect morphologies" of neutral and singly-charged caps are possible at low surface coverage.  These are shown schematically in Fig. \ref{fig5}, and details of the location and existence criterion for scarred states are given in Appendix~\ref{morphologies}.  The diagram in Fig. \ref{fig5} shows the range of existence -- or, the necessary conditions for mechanical equilibrium -- for these states in the $(\Phi,\sigma_B)$ plane: 

\hspace{0.25 cm}

\noindent  I) {\it Neutral caps, dislocation-free} - Defect free caps exist for all coverage and boundary force.

\hspace{0.125 cm}

\noindent II) {\it Charged caps, dislocation-free} - Centered-dislocation caps exist for all coverage and boundary force.

\hspace{0.125 cm}

\noindent III) {\it  Neutral caps, edge-bound scars} - Caps with no excess disclinations and radial scars reaching the outer edge exists over a range of tensile boundary forces $0\leq\sigma_b<Y \Phi/2$.

\hspace{0.125 cm}

\noindent IV) {\it  Neutral caps, center-bound scars} - Caps with no excess disclinations and radial scars extending from the center exists over a range of compressive boundary forces $-Y \Phi/4<\sigma_b\leq0$.

\hspace{0.125 cm}

\noindent V) {\it  Charged caps, edge-bound scars} - Centered-disclination cap with radial scars reaching the outer edge exists over range of compressive boundary force, $-P^c_*(\Phi)<\sigma_b \leq 0$, where $P^c_* =Y(1/12-\Phi/2)$ for $\Phi< 1/18$ and $P^c_*=2^{-3/2}  Y \Phi^{-1/2}/ 27 $ for $\Phi \geq 1/18$. 

\hspace{0.125 cm}

\noindent VI) {\it  Charged caps, center-bound scars} - Centered-disclination cap with radial scars extending from the center exists for all tensile boundary forces, $\sigma_b \geq 0$ and for $\Phi>\Phi_c = 1/12$, for a range of compressive forces $\sigma_b \geq -P_**(\Phi)$, given by eq. (\ref{eq: pss}).

\hspace{0.25 cm}

Table~\ref{tab: morph} summarizes the location of the scarred zones for each morphology, focusing for simplicity in the limiting behavior as $\sigma_b \to0$. We note that for all morphologies, scars cover the entire cap as boundary forces vanishes (that is, $L_1 \to 0$ for edge-bound caps and $L_2 \to W$ for center-bound caps).  That generic fact indicates that the perfect screening of disclination and curvature-induced stresses is possible only for vanishing boundary forces, but for all scarred morphologies.  Despite the apparent degeneracy at $\sigma_b =0$ of all scarred states, the energies of competing morphologies differ in the approach to $\sigma_b \to 0$, a point we return to below.  

Finally, we note the appearance of a critical point $(\Phi_c=1/12,\sigma_b=0)$ for center-bound, charged cap morphology (VI), due to existence of an additional root to eq. \ref{eq: L1L2} for $\Phi>\Phi_c$ for $\sigma_b = 0$. For $\Phi>\Phi_c$, a second stable scar configuration exists with $L_2 < W$ at $\sigma_b =0$.  This solution, which covers only a fraction of the cap will nevertheless have a higher energy that the solution with $L_2=W$ that also exists $\Phi>\Phi_c$.  However, for any finite $\sigma_b$, scarred solutions exist only on the branch with $L_2 < W$, indicating a {\it discontinuous} jump of the scar edge from the boundary for infinitesimal tension for $\Phi>\Phi_c$.  This critical point is located precisely at the point where the disclination charge balances the integrated Gaussian curvature of the cap ($\Phi_c=1/12$).

\begin{table*}
\begin{tabular}{l|c|c|c}
Morphology & $\lim_{\sigma_b \to 0} L_1/W$ & $\lim_{\sigma_b \to 0} (1-L_2/W)$  &  $\lim_{\sigma_{b}\rightarrow 0} \big[ E_{dom}/Y\pi W^2-f(\Phi,\sigma_b) \big]$\\
\hline
\hline
I - neutral, &-& -   &$\frac{\Phi^2}{24}+ \frac{ \sigma_b \Phi}{3Y} $ \\
dislocation-free &&&\\
\hline
II - charged,  & - & -  & $\frac{\Phi^2}{24}-\frac{\Phi}{48}+\frac{1}{288} + \frac{\sigma_b}{Y}\big(\frac{\Phi}{3}+\frac{1}{6}\big)$ \\
dislocation-free & &&  \\
\hline

III - neutral, & $(2 \sigma_b/ \Phi Y)^{1/3} $& 0 & $\frac{\sigma_b^2}{Y^2} \big[\frac{1}{3} \ln \big( \frac{2 \sigma_b}{Y \Phi} \big)-\frac{1}{2}\big]$ \\
edge-bound && & \\ 
\hline

IV - neutral, & $0 $& $\sqrt{|\sigma_b/|Y \Phi}$  & $\frac{4  \Phi^{1/2}}{3} \big(\frac{|\sigma_b|}{Y} \big)^{3/2}$ \\
center-bound && & \\ 
\hline

V - charged,  &$24 |\sigma_b| /Y$&$0$  &$\frac{\sigma_b^2}{Y^2} \ln \big( \frac{24|\sigma_b|}{Y} \big) $ \\
edge-bound & & & \\
\hline

VI - charged,&$0$&$\sqrt{\frac{\sigma_b/Y}{\Phi_c-\Phi}},  \ \Phi < \Phi_c$  & $\frac{4(\Phi_c-\Phi)^{1/2}}{3}\big(\frac{\sigma_b}{Y} \big)^{3/2}, \ \Phi < \Phi_c$   \\
center-bound & &$\frac{\Phi-\Phi_c}{\Phi+\Phi_c/3} +\big(\frac{\sigma_b}{Y}\big)\frac{\Phi+\Phi_c/3}{\Phi-\Phi_c},  \ \Phi > \Phi_c$ & $\frac{4(\Phi-\Phi_c)^5}{3(\Phi+\Phi_c/3)^3}+\big(\frac{\sigma_b}{Y} \big) \frac{4(\Phi-\Phi_c)^3}{(\Phi+\Phi_c/3)}, \ \Phi > \Phi_c$ \\

\end{tabular}
\caption {Summary of predicted behaviors of competing morphologies in the limiting case of vanishing boundary forces.  Energy-densities and scar edge positions include only the leading $\sigma_b$-dependence in this limit.  Here, $\Phi_c=1/12$ corresponds to a bifurcation for the solutions of center-bound, charged scar states, where for $\Phi>\Phi_c$, to solutions are possible, $L_2=W$ and $L_2<W$, where the second branch extends to finite $\sigma_b$. }
\label{tab: morph}
\end{table*}

\section{Continuum limit phase diagram}
\label{continuum}

The energies of competing defect morphologies in the $\epsilon \to 0$ limit derive directly from stress-collapsed solutions, eq. \ref{zone1} - \ref{zone3}, evaluated in the model energy, eq. (\ref{eq:  energy}).  Remarkably, despite the fact the dislocations of finite Burgers vector compose the stress-collapse state and the number of dislocations needed is dependent on $b$, the asymptotic stress and elastic energy itself are {\it independent of lattice spacing}.  The stress-collapsed state describes the ideal relaxation possible via a continuum distribution of dislocations in the $b/W\to 0$ limit (where dislocations are sufficiently abundant to completely collapse hoop stress in the scarred zone).  The  energy of this limiting stress pattern represents the first term (called here the dominant energy, $E_{dom}$) in the asymptotic expansion of the total energy in powers of $\epsilon\propto (b/W)^2$, the parameter which quantifies the dislocation cost relative to the cost of curvature-imposed stretching.  As it is independent of lattice spacing and remains finite as $\epsilon \to 0$, $E_{\rm dom}$ is $O(\epsilon^0)$.  We first restrict our analysis to the dominant energetics, while in a later sections show that finite-$\epsilon$ corrections are essential to resolve the stability of competing defect patterns in the limit of vanishing boundary forces.

The dominant energy, $E_{dom}$, derives from the stress profiles, eqs. (\ref{zone1}-\ref{zone3}), and associated strain and displacement fields.  The total dominant energy can be decomposed as
\begin{equation}
\label{egenI}
E_{dom}=E_I + E_{II}+ E_{III} -2\pi \sigma_b W u^{III}_{r}(r=W)
\end{equation}
where $E_\alpha$ is the elastic energy of zone $\alpha$:
\begin{eqnarray}
E_{I} &=& \frac{\pi}{Y} \int_0^{L_1}dr r~ \Big[(\sigma^I_{rr})^2+(\sigma^I_{\theta \theta})^2- 2 \nu\sigma^I_{rr} \sigma^I_{\theta \theta} \Big] \\
E_{II} &=& \frac{\pi}{Y} \int_{L_1}^{L_2}dr r~ (\sigma^{II}_{rr})^2\\
E_{III}&=& \frac{\pi}{Y} \int_{L_2}^{W}dr r~ \Big[(\sigma^{III}_{rr})^2+(\sigma^{III}_{\theta \theta})^2- 2 \nu\sigma^{III}_{rr} \sigma^{III}_{\theta \theta} \Big] \nonumber\\
\end{eqnarray}
where we use the fact that $\sigma^{I,II,III}_{r \theta}=0$ and $\sigma^{II}_{\theta \theta}=0$.  To determine the work done by boundary forces, we compute the radial displacement of the boundary by integrating $\partial_r u_r = u_{rr} -r^2/(2R^2)$ from the boundary condition $u_r(0) =0$,
\begin{multline}
\label{disg}
u_r^{III}(W)=Y^{-1} \int_0^{L_1} dr~ \big( \sigma^I_{rr}-\nu \sigma^I_{\theta \theta} \big) + Y^{-1} \int_{L_1}^{L_2} dr~  \sigma^{II}_{rr} \\ 
+ Y^{-1} \int_{L_2}^{W} dr~ \big( \sigma_{rr}^{III}-\nu \sigma_{\theta \theta}^{III} \big) -\frac{2 \Phi}{3}W .
 \end{multline}

\begin{figure}
\center\includegraphics[width=0.475\textwidth]{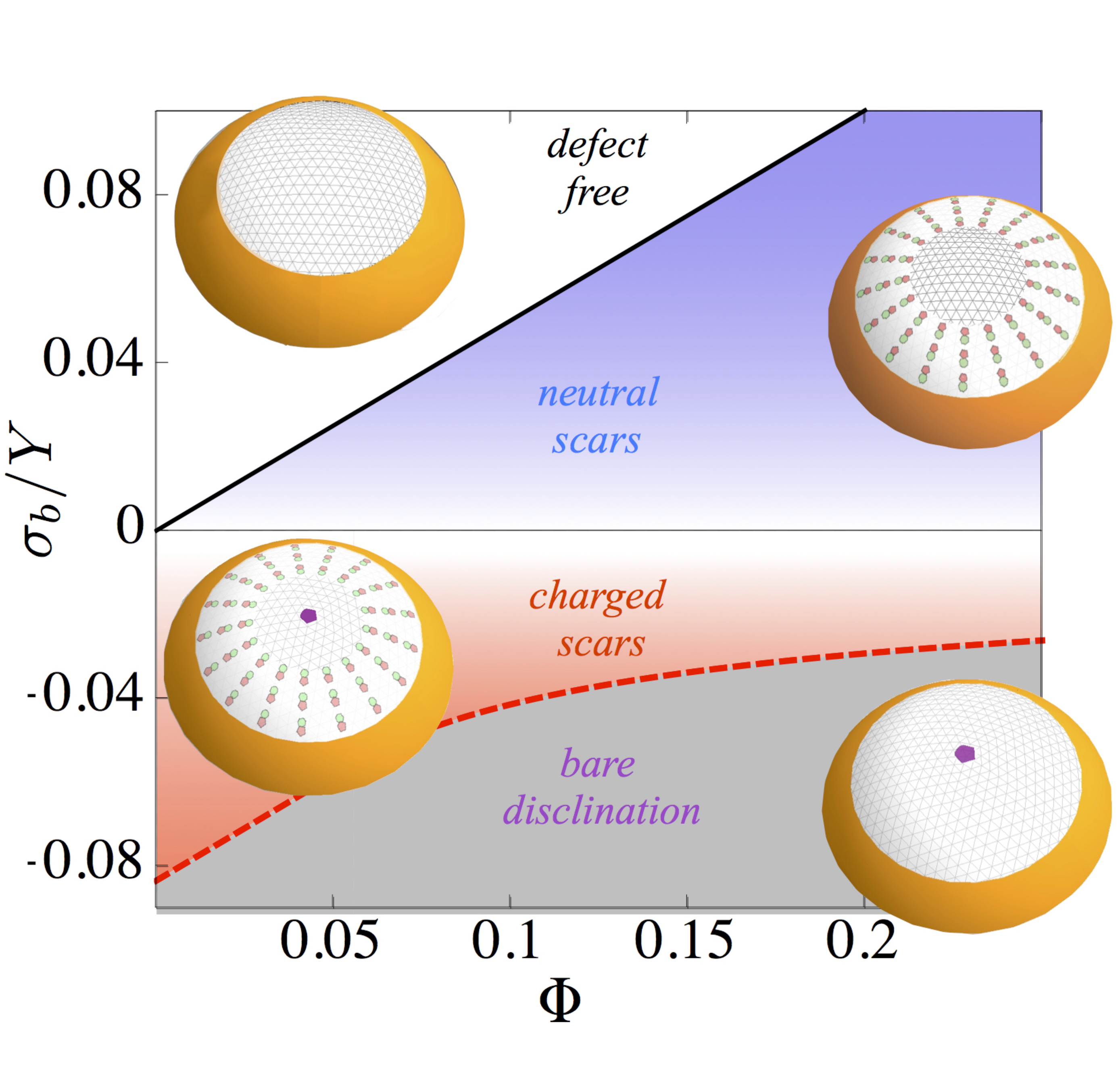}
\caption{Defect phase diagram in the continuum limit where $b/W\to 0$.  Two scarred-morpholgies are favorable for respective tensile and compressive boundary conditions:  neutral, edge-bound scars (state III) and charged, edge-bound scars (state V).  The boundary between these states is the $\sigma_b =0$ where all scarred morphologies achieve total stress collapse through the caps (in the asymptotic $\epsilon \to 0$ limit). }
\label{Fig6}
\end{figure}

Comparing the dominant energies of competing morphologies, we determine the ground state phase diagram for continuum limit, shown in Fig. \ref{Fig6}.  Of the six existing defect morphologies, four appear as minimal energy configurations in the phase diagram. The defect-free cap (disclination and dislocation free) is optimal above the critical tension $\sigma_b > T_*^n=Y \Phi /2$, while the neutral cap with edge-bound scars ($L_2 =W$) is minimal energy between this threshold and the limit of vanishing boundary tension.  Conversely in the tensile regime, the single-disclination, dislocation-free (i.e. $s=0; {\bf b}({\bf x}) =0$) is optimal below the critical compression $\sigma_b < -P_*^c(\Phi)$, given by eq. (\ref{eq: Pc}), that defines the existence of the pattern of edge-bound scars on charged caps.  Between this compressive threshold and the limit of zero boundary compression, charged caps with edge-bound scars are minimal energy.  

In order to understand the generic energetic preference for {\it edge-bound scar} morphologies in the continuum limit phase diagram, we consider competition of neutral and charged scar patterns as $\sigma_b \to 0$.  The generic preference for multi-scar patterns with dislocations extending to the boundary can ultimately be traced to the elastic cost of pulling the ``tips" of center-bound scars from the free boundary into the bulk of the cap $L_2 <W$. This is most transparent in the limit of small $\sigma_b$, where the distance of the scar tips from the cap edge grows as $ \delta \ell = W-L_2 \sim |\sigma_b|^{1/2}$ (see Table ~\ref{tab: morph}) for both neutral and charge center-bound scar states (states IV and VI, respectively)~\footnote{Note that for charged caps with center-bound scars, this scaling holds only for low surface coverage, smaller than the critical value $\Phi_c = 1/12$, while for $\Phi > \Phi_c$, $\delta \ell$ remains finite in the limit $\sigma_b \to 0_+$.}.   As demonstrated in the next section, the dominant cost of scars derives from inter-scar elastic interactions, resulting from elastic strains generated by scar tips in the cap.  For a scar with linear dislocation spacing, $\lambda$, the rotation of the crystallographic directions across the scar, like a low-angle grain boundary, is $\Delta \theta \approx b/\lambda$.~\cite{peach}  At the far-field elastic cost of the scar end, therefore behaves as virtual ``disclination" of charge $\Delta \theta$~\cite{azadi_prl}.  Elastic interactions between disclinations near to a free boundary grow in magnitude as $Y(\Delta \theta)^2   \delta \ell^2$~\cite{romanov} and extend over a distance proportional to $\ell$ due the screening of long-range stresses by the boundary.  Assuming $n_s$ scars in the cap, each tip interactions with roughly $n_s \delta \ell/W$.  Using  eq. (\ref{br}) for the dislocation density $\rho(W)={\bf b}(W)/b$ at the periphery of the cap and $\lambda(W) = 2 \pi W \rho(W)/n_s$, we estimate the total energy of inter-scar interactions as $\lim_{\delta \ell \to 0} E_{\rm inter} \approx Y n_s^2 (\Delta \theta)^2 \delta \ell^2 \sim \delta \ell^3 \sim|\sigma_b|^{3/2}$.  

Due to the isometric radial displacement in the limit of complete stress collapse,  all scarred-morphologies share the same linear dependence on $\sigma_b$ exhibited by $f(\Phi, \sigma_b)$ as $\sigma_b \to 0$.  Therefore, the appearance the $|\sigma_b|^{3/2}$ scaling of scar-scar interactions for center-bound scars implies a higher energy than edge-bound scars, whose distinguishing energetics scale generically as $\sigma_b^2 \ln L_1 \sim \sigma_b^2 \ln |\sigma_b|$ in this limit, which follows from similar arguments for scar tip interactions at the center of the cap.  Hence, we can attribute the generic stability of edge-bound scars over center-bound scars to the anomalously larger elastic cost of pulling scar ``tips" from the free edge, into the bulk of the cap.  Though this argument holds only for the asymptotic approach to $\sigma_b=0$, we find the stability of center-bound scars over edge-bound scars persists over the entire range of finite boundary forces.

\section{Vanishing boundary force and finite lattice spacing}
\label{finiteep}
Because scars are predicted to cover the entire area of both charged and neutral caps when boundary forces vanish, the dominant energies of all scarred patterns, irrespective of total topological charge, are degenerate along the line $\sigma_b=0$.  Thus, resolving the degeneracy between neutral and charged caps the singular limit $\sigma_b \to 0$, requires consideration of the finite-$\epsilon$ contributions to the energies of competing defect patterns.  Here, we show by a scaling argument that the lowest-order corrections to dominant energy, the {\it sub-dominant} energy, can be associated with ``self-energy" of forming scars from multiple, elastically-interacting scars, which is proportional to the total number dislocations in the edge-bound scar patterns (states III and V) , given by eq. (\ref{eq: Ndedge}).  For clarity, we outline the argument for neutral caps ($s=0$), though the argument extends straightforwardly to single-disclination, charged caps (see Appendix~\ref{chargedomsub}).  

Beginning with the elastic energy of the dislocation-free cap $E_{\rm df} \approx Y W^2 \Phi^2$ (assuming $\sigma_b/Y \ll \Phi$ and, hence, $L_1 \ll W$), we consider the change in the cap energy due to presence of multi-dislocation scars.  The relaxation energy $E_{\rm relax}$ is most straightforwardly computed~\cite{lucks, azadi_grason_12} from the mechanical interaction between dislocations and the initially compressive stress at the boundary, proportional to $- Y \Phi$, as edge dislocations of width $b$ virtually ``climbs" from the cap edge a distance $ W-r\approx W$ relaxing energy by
\begin{equation}
E_{\rm relax} \approx - b W Y \Phi N_d  \sim - E_{\rm df} ,
\end{equation}
where we used the fact that  $N_d \sim \epsilon^{-1/2}= \Phi W/b$.  Turning now to elastic interactions between scars, we note again that dominant far-field elastic effect of the scars comes from their ``tips"~\cite{azadi_prl} at $r=L_1$:  For a scar with linear dislocation spacing, $\lambda$, the rotation of the crystallographic directions across the grain-boundary scar is $\Delta \theta \approx b/\lambda$.  At the far-field, the elastic cost of the scar end, therefore behaves as virtual ``disclination" of charge $\sim \Delta \theta$.   Assuming a number of scars $n_s$, the rotation angle of across scars (and the virtual ``charge" of scar tips) is roughly $\Delta \theta \simeq b \lambda \approx b N_d/(n_s W) \approx \Phi/n_s$.   The elastic interaction between central, like-charged virtual disclinations is $~Y (\Delta \theta)^2 W^2$ per pair, leading to a total interaction energy,
\begin{equation}
E_{\rm inter} \approx Yn_s^2(\Delta \theta)^2 W^2 \sim E_{\rm df} .
\end{equation}
Like $E_{\rm relax}$, the scar interactions contribute at the dominant scale, $E_{\rm df}$, and exhibit no dependence on scar number or on the microscopic scale, $b$, identifying these terms with the elastic energy of the dominant stress-collapsed pattern.  The limit of vanishing boundary forces, where the dominant energy vanishes, therefore corresponds to a net relaxation of the elastic energy of initial geometric confinement, as well as the cost of inter-scar repulsions, through scar formation.

Not accounted for in the balance of dominant energies above, is the residual elastic ``self-energy" of scar formation, $E_{\rm self}$, which arise due to the imperfect cancelation geometric stresses for finite $b/W$ (and finite-$\epsilon$).  In addition to the far-field elastic cost of the singular ``tips" of scars, $~Y (\Delta \theta)^2 W^2$ per scar, grain boundaries are characterized by a ``line tension", $\sim Y b^2 \lambda \big[ \ln(\lambda^{-1}/b) + E_c\big]$, where $E_c$ parameterizes the inelastic core energies of dislocations~\cite{peach}, from which we estimate the total self energy of $n_s$ scars
\begin{equation}
\label{eq: self}
E_{\rm self} \approx  \big[ n_s^{-1} + \epsilon^{1/2} \ln (n_s \epsilon^{1/2}) + {\rm const.} \big] E_{\rm df} .
\end{equation}
Optimizing the residual elastic energy of scar $n_s^* \sim \epsilon^{-1/2} \propto N_d$, we find $E_{\rm self} \sim \epsilon^{1/2} $ and that the ratio of this cost relative to the energy of the defect-free state is proportional to $\epsilon^{1/2}\sim b/W $ and hence vanishes as lattice spacing goes to zero. Therefore, we identify the self-energy scar formation, which is, $O(\epsilon^{1/2})$, as the first correction to the $O(\epsilon^{0})$ terms identified as the dominant energy ($E_{\rm relax}$ and $E_{\rm inter}$).  Because differences between these $O(\epsilon^{0})$ terms (which are dominant when $\sigma_b$ is finite and $\epsilon \to 0$) vanish for scarred patterns as for $\sigma_b =0$, the terms which distinguish competing scarred states at zero boundary force derive entirely from these leading finite-$\epsilon$ corrections (which are sub-dominant when $\sigma_b$ is finite and $\epsilon \to 0$) .  

To analyze the case of strictly $\sigma_b=0$ and $\epsilon \ll1$, but finite, we consider the predicted variation of the scar self-energies as estimated above, and compare this picture to numerical simulations of caps along the $\sigma_b=0$ line.  The ratio $E_{\rm self} /N_d \approx Y b^2$ from the analysis above implies the intuitive result that the sub-dominant cost of the scar pattern is simply the $N_d$ times energy cost per dislocation.  Pushing this observation one step further, we propose a heuristic picture where the variation of sub-dominant energy cost, associated with the finite $O(\epsilon^{1/2})$ corrections, is predominantly controlled by variation of the number of dislocations needed to achieve stress-collapse in competing neutral- and charged-cap morphologies.  Along the singular line $\sigma_b =0$, where scars extend through neutral and charged caps, it is straightforward to see that from eq. \ref{eq: Ndedge} that $N_d$ is linear function of $\Phi$ for both states.  Dislocation number increases with $\Phi$ for neutral caps due to the monotonic increase of geometric stress with curvature, while for low surface coverage, $N_d$ decreases with $\Phi$ due to the tendency of curvature to ``neutralize" disclination-induced stresses.   

\begin{figure}
\center\includegraphics[width=0.45\textwidth]{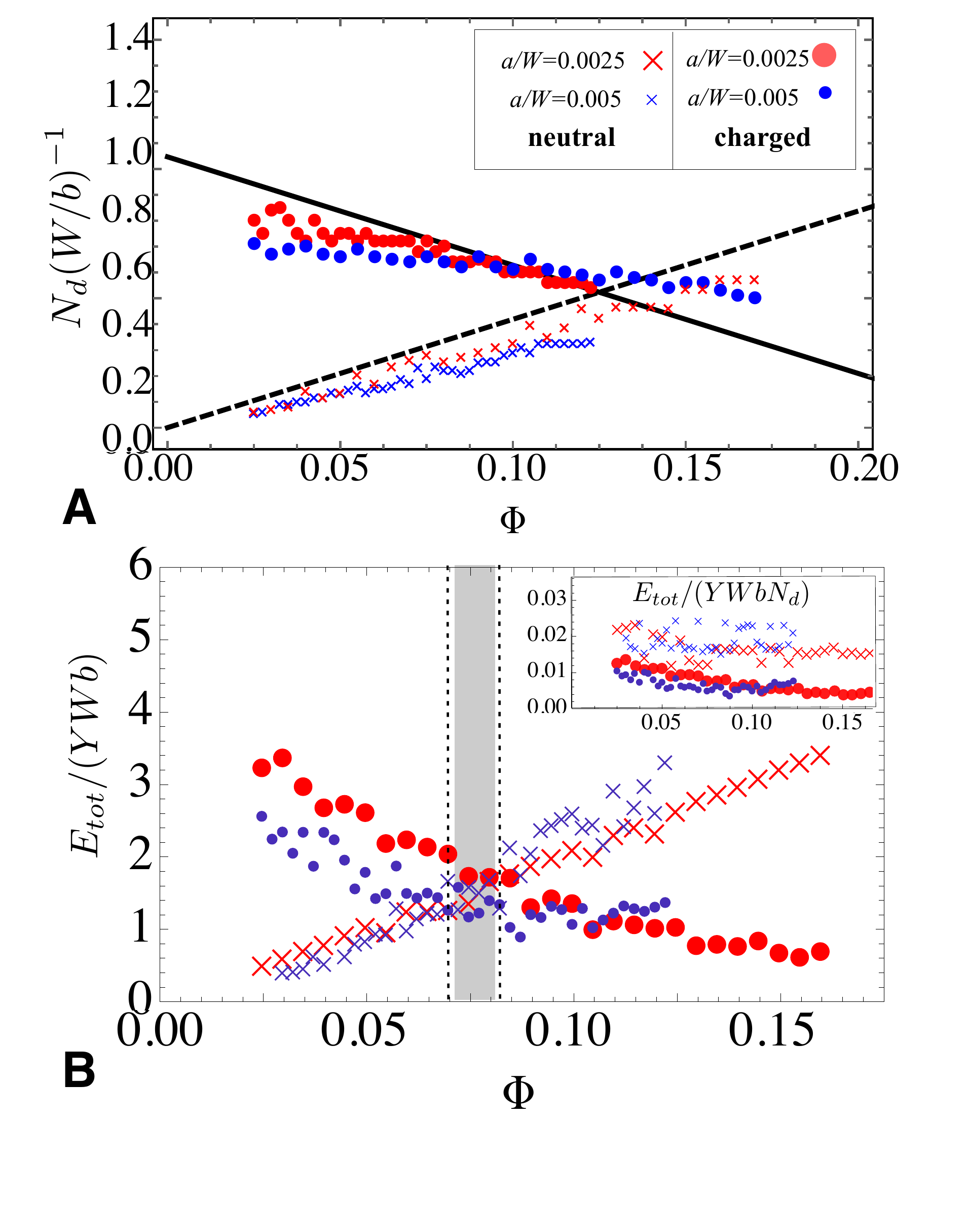}
\caption{Neutral vs. charged scar states for $\sigma_b =0$ for finite, but small, $b/W$.  In (A), we compare the scaled total dislocation number for charged and neutral caps from simulations with two finite lattice spacings, $b/W = 0.0025, 0.005$.  Dashed curves compare predictions from the predictions of the stress-collapse state, eq. (\ref{eq: Ndedge}).  In (B), we compare the normalized total energy of simulated states, indicating a transition from stable neutral scars to charge scars at $\Phi \simeq 0.077\pm 0.005$ (the grey region).  The inset shows the total energy of simulated scar morphologies normalized by $N_d$, indicating a roughly constant energy per dislocation in both morphologies.}
\label{Fig7}
\end{figure}

These linear $\Phi$-dependencies are confirmed in Fig. \ref{Fig7}A where we compare scaled dislocation number in numerically minimized patterns of neutral and charged caps for two ratios of lattice spacing to cap size, $b/W=0.0025$ and $0.005$.  
We show further in Fig. \ref{Fig7}B that the total energies of these competing states crosses from favorable neutral caps to favorable charged caps at value of surface coverage $\Phi \simeq 0.077\pm 0.005$, for both lattice spacings. These numerical results imply that in the asymptotic $\epsilon \to 0$ limit and for strictly zero boundary forces, the presence of multi-dislocation scars reduces the threshold value for stable disclinations from the dislocation-free value of $\Phi=1/6$, to a value remarkably comparable the geometric ``neutrality" condition of $\Phi_c=1/12\simeq0.083$.  In the inset of Fig. \ref{Fig7}B, we compare the total energies per dislocation of the simulated charged- and neutral-scar patterns, showing that the energy per dislocation varies only weakly over the range of $0<\Phi\lesssim 1/6$ in comparison to  relative number of dislocations changes by more than 50\% over the same range.  While the number of dislocations in charged caps is predicted to exceed those in neutral caps up to $\Phi=1/8$, we find that charged caps are stabilized below this surface coverage due to a lower mean energy per dislocation, $\bar{E}_d$:  in charged caps, $\bar{E}_d=0.01 Yb^2$, compared to roughly twice this, $\bar{E}_d=0.02 Yb^2$, for neutral caps.  

Overall, this scaling argument, corroborated by simulations, imply that along the singular line $\sigma_b=0$ the role of surface geometry enters through number of dislocations needed to achieve ``perfect" stress collapse.  In neutral caps, the number of dislocations needed to ``screen"  stresses vanishes as $\Phi \to 0$, while in the presence of the central disinclination requires a finite number of dislocations in flat charged caps ($\propto W/b$).  This balance is altered with increasing surface curvature, which increases $N_d$ for neutral caps due to the increased geometric stress and alternately lowers the number of necessary dislocations in charged caps, due to the tendency of positive curvature to ``neutralize" the disclination stresses.

\section{Discussion}
By exploiting a well-defined asymptotic limit of vanishing lattice spacing, we find that it is possible to compare the morphologically-rich spectrum of multi-dislocation patterns exhibited by crystals bound to curved surfaces.  This theory provides a quantitive framework for the transition from defect-free flat crystals to the ground states of positively-curved crystals, possessing one or more excess five-fold disclinations.  Significantly, we find for zero boundary forces that excess dislocations in the ground state lead to a substantial reduction of the critical curvature (or surface coverage) needed to stabilize to the first excess disclination in the cap.  Moreover, we find that in the asymptotic limit of vanishing lattice spacing, the transition between topologically-neutral to topologically-charged defect patterns becomes relatively insensitive to surface curvature, exhibiting instead a very sharp (discontinuous as $\epsilon \to 0$) dependence on forces at the boundary of cap.  We note, in passing, that identical predictions hold for crystalline sheets on a constant-negative curvature surface provided we simply invert the  sign of surface curvature (or $\Phi$), disclination charge and boundary forces in the analysis of the cap.  

A key point of comparison of this theory are the experiments of Irvine {\it et al.} which studied optimal defect patterns formed by charge colloids on oil-water droplets, where the surface coverage was varied through changing the contact angle with partially wetting surfaces.  In these experiments the ratio crystal dimension to lattice spacing is far from the continuum limit ($W/b \sim 10-20$) and the boundary conditions on the crystal at the drop edge are not precisely defined, yet observed defect patterns  are clearly characterized by large scars, suggesting that multi-dislocation patterns play an important role in stabilizing excess disclinations. Indeed, these experiments find that the transition from $S=0$ to $S \neq0$ seems to fall along topologically-charged neutrality line, consistent with our prediction for the stability condition of the $S=\pi/3$ at $\sigma_b=0$, as opposed to the geometric ``overcharging" expected in the absence of multi-dislocation scars. 

Before concluding, we discuss two further open issues regarding the optimal structural response of crystals confined to spherical surfaces.  The first concerns the crossover from the defect phase diagram for the sheet is comparable to the lattice dimension, $W \gtrsim b$ (Fig.~\ref{Fig1}) to the asymptotic regime of the continuum limit where $W \gg b$ (Fig.~\ref{Fig6}) .  Our results suggest that as the cap size $W/b$ is increased from order unity to $\gg 1$ (corresponding to $\epsilon$ decreasing from order $\Phi^{-1}\gtrsim1$ to $\ll 1$) the phase boundary $\sigma^*(\Phi)$ flattens from a linear function of $\Phi$ (neutral/charged boundary  controlled by both curvature and external force) to the $\sigma_b=0$ line (transitioned controlled by boundary force only), with a critical surface coverage shifting from $\Phi=1/6$ down to roughly $\Phi\simeq1/12$.  In this study, we are able to address either the limit where boundary forces are finite and $b/W \to 0$ (in which dominant energy scales are $O(\epsilon^0)$) , or instead where boundary forces vanish, while $b/W$ remains small but finite (in which dominant energy scales are $O(\epsilon^{-1/2})$).  Determining the behavior of $\sigma^*(\Phi)$ for small but finite-$\epsilon$ needed to capture the crossover between the large- and small-lattice spacing regimes requires comparison of the competing charged- and neutral-scar patterns when the $O(\epsilon^0)$ and $O(\epsilon^{1/2})$ termsare necessarily comparable.  In this case, the separation of energy scales cannot be predicated on the limiting assumption of prefect stress collapse, which was assumed to optimize the (dominant) elastic energy in the $\epsilon \to 0$ limit.  For finite $b$ optimal states will achieve a slightly different partition of energy between dislocation relaxation energy, scar interactions and self-energy than was outlined above in Sec.~\ref{finiteep}.  One way to see this is from the balance of the radial Peach-Koehler force $ \epsilon_{ij} \sigma_{jk} b_k$ and the radial self-interaction force of dislocations in a circular crystal~\cite{azadi_grason_12},  $f_{\rm self} (r) = (Yb^2/4 \pi) r^3/(W^2-r^2)$, which suggest that mechanical equilibrium requires $\sigma_{\theta \theta} = -f_{\rm self} (r)/b \propto b $ in the scarred zone, rather than the strict assumption perfect stress-collapse (or $\sigma_{\theta \theta} =0$).  The degree of this ``imperfect stress collapse" is dependent on lattice dimension, implying $\epsilon$-dependent corrections to the ``dominant" elastic energy computed from the perfect-stress collapse condition invoked here.  It is likely such $b/W$ corrections to the ``dominant" elastic energy will play an important role in the evolution of the charge-to-neutral boundary away from the $\epsilon \to 0$ limit.

A second important open question involves the competition of defect-mediated relaxation of curvature-imposed frustration, with other structure modes possible for a self-assembled, sheet-like crystal.  Here, we find that when $b/W \gg 1$ excess dislocations  greatly reduce the cost of geometric frustration and fundamentally alter the transition between topologically neutral and charge defect patterns.  This is, of course, predicated on the assumption that crystal boundaries remain roughly circular and the shape of the crystals conforms perfectly to the imposed spherical shape.  Previous theories~\cite{schneider, bruinsma} and recent experiments of colloidal crystals on curved droplets~\cite{manoharan} show that when inter-element forces are sufficiently brittle and line-tension of the crystal is sufficiently low, the optimal state avoids topological defects through reshaping the boundary of the crystal to an anisotropic, ribbon-like morphology.  Moreover, in a previous study we have addressed the possibility of a wrinkle-to-scar transition possible when a crystalline cap forms on substrate is sufficiently deformable to allow for both out-of-plane elastic deformation (e.g. wrinkles, folds) and in-plane, ``plastic" deformation (topological defects).  This  study compared wrinkle  vs. neutral scar patterns on crystalline caps under adhesive tension.  In this regime of parameter space, we found that the degree of elastic relaxation possible via dislocation scars in the $b \to 0$ limit is identical to what can be achieved by highly-wrinkled states for infinitely-bendable sheets.  This connection between the principles of defect-mediated stress collapse and the so-called ``tension field theory" of infinitely-bendable sheets~\cite{stein, pipkin, schroll} implied that the relative stability of wrinkled or scarred states of the cap is entirely dictated by microscopic energetics associated with subdominant costs for wrinkles or defects, since their dominant elastic energies are identical.  

In the present case, where we generalize the problem to include both the possibility of excess 5-fold disclinations and compressive boundary stresses, we expect a significant departure from the ``near degeneracy" of wrinkles and scars predicted under tensile boundary forces.  While out-of-plane deformation of sheets is only able to collapse compression (through absorbing excess length), dislocations are able to collapse both compression {\it or} tension:  flipping the orientation of dislocation changes a removed partial row of sites to an added partial row.  Indeed, we find the existence of defect morphologies which collapse tension rather than compression via the reversal of the Burgers vector along the hoop direction.  One of these states, edge-bound, charged scars (V) becomes energetically favored for $\sigma_b <0$.  This mechanical state, which sustains compression but excludes tension throughout the cap, is not achievable via out-of-plane elastic deformation (such as, wrinkling).  This implies that compressive boundary forces (and, possibly, also negative curvature) break the degeneracy between ``stress-collapse" states of wrinkled and scarred states of caps, leading to the selection of cap morphology deriving from the distinction between the elastic energy relaxation achieved via defect-mediated vs. wrinkle-mediated response.  

\section{Conclusion}

Beyond a new understanding of structure formation in existing materials like particle-coated droplets, our framework for emergent defect patterns introduces new directions for 2D materials design. For instance, recent theoretical and experimental work on ``designer graphene" indicate that mechanical properties of atomically-thin sheets can be dramatically remodeled by the presence of controlled amounts of dislocations \cite{GR1,GR2,GR3}.  In this light, our framework may be exploited to direct the number, location, orientation and pattern of defects on a crystalline sheet through manipulation of a combination of boundary forces and geometry of the sheet, which in turn may lead to new paths for engineering two-dimensional with spatially heterogeneous metamaterial properties.

\begin{acknowledgments}
We are grateful to B. Davidovitch numerous discussions regarding this research.  Acknowledgement is made to the NSF (CAREER grant no. DMR 09-55760), the Donors of the American Chemical Society Petroleum Research Fund (ACS PRF no. 54513-ND) and the Alfred P. Sloan Foundation for support of this research.\end{acknowledgments}

\appendix

\section{Stress-collapsed morphologies}

\label{morphologies}
In the following subsections, we summarize the structure and energetics of competing radial scar morphologies.  Consistent with numerical simulation results (Fig.~\ref{Fig4}), we consider, separately the cases of edge-bound and center-bound scars, by analyzing the equation of state, eq. ~\ref{eq: L1L2}), for the respective cases of $L_1\geq 0; L_2 = W$ and $L_1 =0; L_2 \leq W$.  Additionally, we consider, alternately, neutral and charged morphologies, corresponding to the respective $s=0$ and $s =\pi/3$ cases of the stress-collapsed solutions.  For each of these four competing morphologies --- neutral/charged and edge-bound/center-bound scars --- we summarize predictions for the positions of scarred-zones and dominant energetics.  The limiting behavior of these morphologies, as well as the competing dislocation-free states, is summarized in Table~\ref{tab: morph}.  Note that we follow the morphology labels (I-VI) defined in Fig.~\ref{fig5}.

\subsection{Neutral caps, edge-bound scars (morphology III)}
Here we consider the case $s=0$ and $L_2 = W$ for which the equation of state, eq. (~\ref{eq: L1L2}), reduces to the simple condition,
\begin{equation}
\sigma_b/Y = \frac{L_1^3}{8 R^2} ,
\end{equation}
which has the solution for the inner-edge of the scarred zone
\begin{equation}
L_1 = W (\sigma_b/T^n_*)^{1/3}
\end{equation}
with $T^n_* \equiv Y \Phi/2$.  In the limit of vanishing tensile stress, scar edges exhibit a singular approach to the center $\lim_{\sigma_b \to 0_+} L_1 \propto \sigma_b^{1/3}$.  The requirement that $L_1\geq 0$ and $L_1<W$ define the existence criteria for edge-bound, neutral scars:  $0\leq \sigma_b <T^n_*$.  

The dominant energy of edge-bound, neutral scars is given by 
\begin{equation}
\label{edomN}
\frac{E_{dom, III} }{\pi W^2} =\frac{\sigma_b^2}{Y}\Big[\frac{1}{3}\ln\left(\frac{\sigma_b}{T_*^n}\right)-\frac{1}{2} \bigg]+f(\Phi,\sigma_b) ,
\end{equation}
where $f(\Phi,\sigma_b) \equiv -\sigma_b^2 (1-\nu)/Y + 4 \Phi \sigma_b /3$ is a morphology-independent term in the free-energy density of all stress-collapsed states.

\subsection{Neutral caps, center-bound scars (morphology IV)}
Here, we consider the case $s=0$ and $L_1 = 0$ for which the equation of state, eq. (~\ref{eq: L1L2}), reduces to the form,
\begin{equation}
\label{eq: eosIV}
\sigma_b/Y =-  \frac{(W^2-L_2^2)^2}{16 W^2 R^2} ,
\end{equation}
which has the solution for the outer-edge of the scarred zone
\begin{equation}
L_2 = W \sqrt{1-(-\sigma_b/P^n_*)^{1/2}} 
\end{equation}
with $P^n_* \equiv Y \Phi/4$.  In this state, scars approach the edge of the cap in the limit of vanishing compression, as $\lim_{\sigma_b \to 0_-} (W-L_2) \propto |\sigma_b|^{1/2}$.  The range of real solutions to eq.~(\ref{eq: eosIV}) with $W\geq L_2>0$ define the existence criteria for center-bound, neutral scars:  $0\geq \sigma_b > -P^n_*$.  

The dominant energy of this state is given by 
\begin{multline}
\label{edomIV}
\frac{E_{dom, IV} }{\pi W^2}=-\frac{4 \Phi \sigma_b}{3} \Big(1-(-\sigma_b/P_*^n)^{1/2} \Big)\\ \times\Big(1-\sqrt{1- (-\sigma_b/P_*^n)^{1/2}}\Big)  +f(\Phi,\sigma_b) 
 \end{multline}
In the limit of weak compression, the most significant contribution to the energy difference with competing morphologies derives from the first term, which scales as $\sim|\sigma_b|^{3/2}$.  The physical origins and consequences of this singular dependence on boundary forces are discussed in Sec.~\ref{continuum}.

\subsection{Charged caps, edge-bound scars (morphology V)}
Here, we consider the case of a centered, 5-fold disclinatiion, $s=\pi/3$ and scars extending to the cap edge $L_2= W$ for which the equation of state,  eq. (\ref{eq: L1L2}), reduces to the cubic form,
\begin{equation}
\label{eq: eosV}
\sigma_b/Y =-\frac{L_1}{8W} \Big(\frac{1}{3}-\frac{2 \Phi L_1^2}{W^2}  \Big) .
\end{equation}
Under weak boundary compression, $\sigma_b<0$, this equation can have two real roots for $L_1$, the smaller of which corresponds to the lower-energy branch, with scars covering a larger fraction of the cap.  For example, in the limit of weak compressive forces, the inner edge of the scars approaches the center as 
\begin{equation}
\lim_{\sigma_b \to 0_-} L_1/W = - \frac{24  \sigma_b}{Y}\Big[1+216 \Phi(\sigma_b/Y)^2\Big],
\end{equation}
which also shows that tendency of the scar edge to pull away from the cap center with increasing curvature.  The existence of the edge-bound, charged cap is constrained by two conditions:  1) the existence of real value solutions for $L_1$, which requires $\sigma_b \geq -2^{-3/2}Y \Phi^{-1/2}/27$ and 2) the requirement that scar edges are within the cap, $L_1 \leq W$, which requires that $\sigma_b \geq - Y(1/6-\Phi)/2$.  These two conditions meet at a value of cap curvature $\Phi=1/18$, such that we may define the existence of the charged-cap, edge-bound scar state as the range $0\geq \sigma_b>-P_*^c$ where
\begin{equation}
\label{eq: Pc}
P_*^c = \left\{ \begin{array}{ll} Y(1/6-\Phi)/2, & {\rm for} \Phi < 1/18  \\  \\ 2^{-3/2}Y \Phi^{-1/2}/27, & {\rm for} \ \Phi \geq 1/18  \end{array} \right.
\end{equation}

For a solution of eq. (\ref{eq: eosV}) for $L_1$ the dominant energy of this state is
\begin{multline}
\label{edomV}
\frac{E_{dom, V} }{\pi W^2}=\frac{YL_1^2}{W^2}\Big[\frac{\Phi^2 L_1^4}{24 W^4} -\frac{\Phi L_1^2}{48 W^2} + \frac{1}{288} \Big]-\frac{\sigma_b^2}{Y}\ln(W/L_1)\\
+\frac{\sigma_b L_1}{W}\Big(\frac{1}{12}-\frac{\Phi L_1^2}{8 W^2} \Big)+f(\Phi,\sigma_b) 
 \end{multline}
Here, we note that the distinguishing costs of this morphology vanish smoothly with vanishing boundary stress, as $\sim \sigma_b^2\ln \sigma_b$, deriving from the energy of pulling scar ends from cap center.  

\subsection{Charged caps, center-bound scars (morphology VI)}
Here we consider $s=\pi/3$ and $L_1=0$ for the case of scars bound to the center of a charged caps.  The from these conditions, the equation of state  equation of state, eq. \ref{eq: L1L2}), has the form
\begin{equation}
\label{eq: eosVI}
\frac{\sigma_b}{Y} = -\frac{1}{12} \Big[ \ln(L_2/W)+ \frac{W^2-L_2^2}{2W^2}\Big]-\Phi\frac{(W^2-L_2^2)^2}{4W^4}
\end{equation}
This equation has real solutions for $L_2 \leq W$ all tensile boundary forces, with an $L_2$ that generically decreases with $\sigma_b$.  For example, for extremem limit of large tensions ($\sigma_b \gg Y/12$) the solution to eq. (\ref{eq: eosVI}) has the solution $L_2 \simeq W\exp\big[-12 \sigma_b/Y\big]$, indicating a vanishing, yet finite, scarred zone at the center of the cap.  For the case of vanishing boundary forces there is always root $L_2 =W$, corresponding to scars covering the entire cap.  To consider the approach as $\sigma_b \to 0_+$, we define $L_2 = W - \delta \ell$ and expand eq. (\ref{eq: eosVI}) for $\delta/W \ll1$ yeilding an approximate equation of state,
\begin{equation}
\label{eq: eosVIms}
\frac{\sigma_b}{Y} \simeq \Big(\frac{1}{12}- \Phi\Big) (\delta \ell/W)^2 + \Big(\frac{1}{36}+ \Phi\Big)  (\delta \ell/W)^3,   \ \ \ {\rm for} \ \delta \ell/W \ll1 
\end{equation}
The point at which the quadratic coefficient changes sign defines a critical surface coverage, $\Phi_c = 1/12$, which also corresponds to the point where the integrated curvature of the cap perfectly neutralized the 5-fold defect.  The limiting behavior of the scar edge as $\sigma_b \to0$ therefore splits into three regions
\begin{equation}
\label{eq: deltaell}
\lim_{\sigma_b \to 0_+} \delta \ell = W \times \left\{ \begin{array}{ll} \sqrt{\frac{ \sigma_b/Y}{\Phi_c - \Phi} } & {\rm for} \ \Phi< \Phi_c \\ \\ 
\Big(3 \sigma_b/Y)^{1/3} &  {\rm for} \ \Phi= \Phi_c \\ \\ 
\frac{\Phi-\Phi_c}{\Phi+ 1/36}+\frac{\Phi+ 1/36}{\Phi-\Phi_c}(\sigma_b/Y) & {\rm for} \ \Phi> \Phi_c \end{array} \right.
\end{equation}
The existence of this state extends over the full tensile regime $\sigma_b >0$, and somewhat into the compressive regime for large curvature, when $\Phi> \Phi_c$.  In terms of the critical compression
\begin{equation}
\label{eq: pss}
P_{**}^c = Y  \left\{ \begin{array}{ll} 0  & {\rm for} \ \Phi< \Phi_c \\  \frac{\Phi}{4} \Big(1- \frac{\Phi_c^2}{\Phi^2} \Big) +\frac{1}{24} \ln (\Phi_c/\Phi) & {\rm for} \ \Phi> \Phi_c \end{array} \right.
\end{equation} 
center-bound scars exists on charged caps when $\sigma_b \geq P_{**}^c$.


It terms of a solution for the scar edge, $L_2$, the dominant energy of the center-bound, charged scars state is

\begin{multline}
\label{edomVI}
\frac{E_{dom, VI} }{\pi W^2}=\frac{Y (W^2-L_2^2)^2}{W^4}\bigg[\frac{\Phi^2 }{24}\Big(1+2 \frac{L_2^2}{W^2}+3 \frac{L_2^4}{W^4}\Big)\\ - \frac{\Phi}{48}\Big(1+2 \frac{L_2^2}{W^2} \Big) + \frac{1}{288} \bigg] +\frac{\sigma_b}{3 Y}\bigg[\Big(\frac{1}{2} - \Phi\Big) \\ - \Big(1-4 \Phi \frac{L_2^2}{W^2} \Big)\frac{ L_2}{W} \bigg]+f(\Phi,\sigma_b) 
 \end{multline}
The energetics simply considerably in the limit of $W - L_2 =\ell \to 0$, where by using the limiting form of the equation of state, eq. (\ref{eq: eosVIms}), we have 
\begin{equation}
\frac{\lim_{\delta \ell \to 0} E_{dom, VI} }{\pi W^2} = \frac{4 (\Phi-\Phi_c)^2}{3} \delta \ell^3 +f(\Phi,\sigma_b) ,
\end{equation}
where the dependence on boundary forces derives from the $\sigma_b$-dependence of $\delta \ell$ described by eq. (\ref{eq: deltaell}).  The $\sim \delta \ell^3$ dependence exhibited here is common to the center-bound scar morphologies, both charged and neutral.  The origin and consequence of this term are discussed in Section \ref{finiteep}.

\section{Discrete dislocation simulations}
\label{simulations}
Here, we briefly summarize the method for numerically simulating the elastic energy of multi-scar patterns in neutral and charged caps.  This method has been applied and described previously in refs. \cite{azadi_grason_12} and \cite{azadi_prl}.   The elastic effects of discrete dislocations are modeled from the solutions to the compatibility equation, main text eq. (2), for dipolar sources to the bi-harmonic equation for the Airy stress.  From this equation, the stresses generated by a dislocation with Burgers vector, ${\bf b}_\alpha$ at position ${\bf r}_\alpha$ may be computed, and in turn, the elastic interactions of the dislocation with itself, other dislocations and other sources of stress in the system (i.e. disclinations, boundary forces, curvature) may be derived~\cite{azadi_grason_12}.  From these expressions, we compute the energy of a cap possessing an arbitrary array of $N_d$ dislocations as
\begin{multline}
E_{tot}= E_{\rm df}+\sum_{\alpha=1}^{N_{d}}\big[E^D_{self} ({\bf b}_\alpha, {\bf r}_\alpha)+E^D_{relax} ({\bf b}_\alpha, {\bf r}_\alpha) \big] \\ + \sum_{\alpha=1}^{N_{d}}\sum_{\beta<\alpha}^{N_{d}}E^D_{int}  ({\bf b}_\alpha, {\bf r}_\alpha; {\bf b}_\beta, {\bf r}_\beta) ,
\label{total}
\end{multline}
where $E_{\rm df}$ is the energy of the dislocation-free state, SI eq. (\ref{eq: energy}).  The expressions for elastic self-interaction of a dislocation, $E^D_{self} ({\bf b}_\alpha, {\bf r}_\alpha$ and the pair-wise elastic interaction between dislocations, $E^D_{int}  ({\bf b}_\alpha, {\bf r}_\alpha; {\bf b}_\beta, {\bf r}_\beta) $, are given in ref. \cite{azadi_prl} (Supporting Material).  The coupling of dislocations to the stresses generated by the cap curvature, boundary stresses and the possible central disclination are encoded in $E^D_{relax} ({\bf b}_\alpha, {\bf r}_\alpha)$, which is derived from the work done by the Peach-Koehler force on the dislocation from ``non-dislocation" stresses, $\sigma_{ij}^{\rm df}(r)$, when pulling it along a radial path from the free-boundary into the bulk of the cap,
\begin{equation}
E^D_{relax}({\bf b}, r)= b_\theta \int_{r}^{W} dr'  \sigma^{\rm df}_{\theta \theta}(r') ,
\end{equation}
where we used the fact that $\sigma^{\rm df}_{r \theta}=0$, while the dependence of hoop stresses on curvature, boundary forces and the central disclination are given in SI eq. (\ref{stress_5fold}).  From this expression, it is clear that the maximal energy relaxation is acheived for dislocations polarized along the azimuthal direction, corresponding to the addition/removal of a partial row of lattice positions extending from the dislocation to the cap edge along the radial direction.  Hence, in our simulated ground states, we assume that ${\bf b}_\alpha = b \hat{\theta}$ for all dislocations.

Given this discrete-dislocation energy, we simulate the structure and energetics of ``$n$-fold" patterns of multi-dislocation scars possessing total dislocation number $N_{d}$ for a given $W/b$ ratio and surface coverage $\Phi$, boundary stress $\sigma/Y$ and central disclination charge.   In this symmetry, dislocations are constrained to $n_{s}$ identical radial lines (scars), equally spaced at angular intervals of $2\pi/n_{s}$ on the cap.  It was shown previously~\cite{azadi_prl} that $n$-fold symmetric scars emerge spontanously as ground states of neutral caps provided their length is not too long.  Even in cases where optimal scars are not $n$-fold symmetric, the dominant energetics and scaling dependences of the multi-scar ground state are well-modeled by the behavior $n$-fold symmetric patterns.  The radial positions of the $N_d/n_s=M$ concentric rings (constrained to an integer) of dislocations dislocations are initialized randomly, then relaxed via steepest descent according the discrete-dislocation energy.  After relaxation of position, the procedure is repeated, varying the scar number to find the optimal $n_{s}$ for a given $N_d$, $\sigma$, $b/W$ and  $\Phi$. We compute the elastic energy of the simulated system, $E_{tot} (N_d; \sigma, b/W, \Phi, s)$ for a range of possible dislocation numbers, $N_{d}=N^{c}_{d}\pm 0.35 N^{c}_{d}$, where we use the prediction of $N^{c}_{d}$ from continuum theory, SI eqs. (\ref{eq: Ndedge}) and (\ref{eq: Ndcenter}), as the initial guess.  From this set of optimized state, we select $N_d$ corresponding to the lowest energy state.  

For the results shown in main text Figures 3 and 4, discrete dislocation simulations were carried out over a  curvatures: $\Phi=0...0.17$, for two different ratios of lattice spacing to cap size, $b/W=0.0025, 0.005$. 

\section{Dominant vs. sub-dominant scar energetics for charged caps}

\label{chargedomsub}
Here, we briefly overview the scaling analysis of the multi-scar, charged cap morphologies to classify the energetic contributions of the state into dominant (finite as $b/W \to 0$) and subdominant (vanishing as $b/W \to 0$).  We follow the same arguments presented in Sec.~\ref{finiteep} for the neutral scar morphology again focusing on the limit $\sigma_b \to 0$, and show how these generalize in the presence of a central disclination.  For the charged cap, the energy of dislocation-free cap can be written as,
\begin{multline}
\lim_{\sigma_b \to 0} E_{\rm df}(s=\pi/3)  = Y \pi W^2   \Big(\frac{\Phi^2}{24}-\frac{\Phi}{12} +\frac{1}{288}\Big) \\ \sim Y W^2 f_{E}( \Phi) ,
\end{multline}
where $f_{E}( \Phi)$ is a dimensionless function  remains finite as $\Phi \to 0$.  Likewise, stress generate by the combination of the disclination and curvature, from eq. (\ref{stress_5fold}), can be shown to be proportional to $Y$ times a dimensionless function of $\Phi$.  Therefore, the relaxation energy achieved by pulling $N_d$ dislocations into the charged caps has a similar form as main text eq. (11) for the neutral cap (up to dimensionless function of $\Phi$),
\begin{equation}
E_{\rm relax} (s=\pi/3) \approx - b W Y  N_d  \sim - E_{\rm df} f_r (\Phi).
\end{equation}
where from eq. (\ref{eq: Ndedge}) we have used that $N_d\sim W/b $ and $f_r(\Phi)$ is a dimensionless function that remains finite as $\Phi \to 0$.  Thus, the presence of the central disclination, and the interaction between dislocations in scars with that central disclination do not alter the conclusion that $E_{\rm relax} \sim - E_{\rm df}$ and relaxation energy of dislocation scars remains finite as $b\to 0$.  

The scaling of the energies for interaction between scars, $E_{\rm inter}$, and the elastic self energy of scar formation, $E_{\rm self}$, follows the identical argument as presented in the main text for neutral scars, only modified according to $N_d= W/b f_{N_{d}} (\Phi)$, where again $f_{N_d} (\Phi)$ is a dimensionless function that remains finite as $\Phi \to 0$.  Hence, it is straight forward to show that the central results that $E_{\rm inter} \sim E_{\rm df}$ and $E_{\rm self} \sim E_{\rm df}/n_s^*$, where the optimal scar number $n_s^* \sim W/b$ diverges in the continuum limit.  Thus, like the case of the neutral cap, $E_{\rm relax}$ and $E_{\rm inter}$, remain finite as $b \to 0$ and can therefore be associated with the dominant energetics captured by the elastic free energy of the stress-collapse solutions, while the self-energy of scar formation $E_{\rm self}\approx Y b^2 N_d \sim b/W$ can be associated with the sub-dominant correction to the limit of perfect stress collapse only possible with $b/W$ is strictly 0.

\end{document}